\documentclass{article}

\usepackage[english]{babel}
\usepackage{authblk}
\usepackage[letterpaper,top=2cm,bottom=2cm,left=3cm,right=3cm,marginparwidth=1.75cm]{geometry}
\usepackage{adjustbox}
\usepackage{amsmath}
\usepackage{graphicx}
\usepackage[colorlinks=true, allcolors=blue]{hyperref}
\usepackage{wasysym}
\usepackage{subfig}

\usepackage{floatrow}
\newfloatcommand{capbtabbox}{table}[][\FBwidth]

\usepackage[numbers,sort&compress]{natbib}

\title{X-ray phase-contrast micro tomography of soft tissues using a compact laboratory system with two-directional sensitivity}
\author[1]{Carlos Navarrete-Le\'{o}n}
\author[1]{Adam Doherty}
\author[1]{Savvas Savvidis}
\author[2,3]{Mattia F. M. Gerli}
\author[4]{Giovanni Piredda}
\author[1]{Alberto Astolfo}
\author[5,1]{David Bate}
\author[1]{Silvia Cipiccia}
\author[1]{Charlotte K. Hagen}
\author[1]{Alessandro Olivo}
\author[1]{Marco Endrizzi}

\affil[1]{Department of Medical Physics and Biomedical Engineering. University College London, Malet Place, Gower Street, WC1E 6BT London, United Kingdom}
\affil[2]{UCL Division of Surgery and Interventional Science, Royal Free Hospital, London, United Kingdom}
\affil[3]{Stem Cell and Regenerative Medicine Section, Great Ormond Street Institute of Child Health, University College London, WC1N 1EH, United Kingdom}
\affil[4]{Research Center for Microtechnology, Vorarlberg University of Applied Sciences, Hochschulstr. 1, 6850, Dornbirn, Austria}
\affil[5]{Nikon X-Tek Systems Ltd, Tring, Herts, HP23 4JX, United Kingdom}

\date{}

\begin{document}
\maketitle

\begin{abstract}
X-ray micro tomography is a non-destructive, three-dimensional inspection technique applied across a vast range of fields and disciplines, ranging from research to industrial, encompassing engineering, biology and medical research. Phase-contrast imaging extends the domain of application of X-ray micro tomography to classes of samples that exhibit weak attenuation, thus appear with poor contrast in standard X-ray imaging. Notable examples are low-atomic-number materials, like carbon-fibre composites, soft matter and biological soft tissues. We report on a compact and cost effective system for X-ray phase-contrast micro tomography. The system features high sensitivity to phase gradients and high resolution, requires a low-power sealed X-ray tube, a single optical element, and fits in a small footprint. It is compatible with standard X-ray detector technologies: single-photon-counting offers higher sensitivity whereas flat-panels are preferred for a larger field of view. The system is benchmarked against known-material phantoms and its potential for soft-tissue three-dimensional imaging is demonstrated on small-animal organs: a piglet oesophagus and a rat heart.
\end{abstract}

\section*{Introduction}

X-ray micro tomography has become an invaluable tool for the non-destructive volumetric imaging of samples for a wide variety of fields, ranging from medical sciences to metrology and manufacturing \cite{Withers2021,Stock2008,Maire2013}. Despite continuous progress in X-ray micro tomography technology, three dimensional imaging of low-density samples, such as soft tissue specimens, remains challenging due to the weak X-ray attenuation contrast which is linked to low atomic number materials. X-ray phase-contrast micro tomography (XPC$\mu$T) can overcome this limitation by exploiting also the phase-shift experienced by the wave when traversing the object for generating image contrast \cite{Paganin2006}, thus extending the applicability of micro tomography to a broader range of samples and disciplines.

The technology for X-ray phase-contrast imaging is more demanding in comparison to what is needed for conventional X-ray imaging, specifically in terms of the required spatial and temporal coherence of the radiation \cite{Paganin2006}. Whilst synchrotron radiation facilities can achieve high degrees of coherence, that is not often the case for laboratory-based systems, where the power density for X-ray generation is limited and a small footprint is required. In turn, this means that one has to choose between having a small or a powerful source, and working with monochromatic radiation often requires long exposure times. Many solutions have been identified and pursued over the past twenty five years to overcome these limitations and facilitate the translation of X-ray phase-contrast imaging to a laboratory setting. Without compiling a complete history, here we summarise some of the milestones: microfocus source and crystal combination \cite{Davis1995}, free space propagation with polychromatic radiation \cite{Wilkins1996}, Talbot-Lau interferometry \cite{Pfeiffer2006,Bech2009,Herzen2009}, Edge Illumination \cite{Olivo2007,Hagen2014}, Fourier analysis \cite{Wen2010}, liquid metal jet sources \cite{Hemberg2003}, universal Moir\'e effect \cite{Miao2016}, speckle-based imaging also in combination with liquid metal jet sources \cite{Zanette2014,Zanette2015,Zhou2018,Wang2016} and implicit tracking \cite{Quenot2021}; we refer to some excellent reviews on the topic for a more comprehensive list and discussion \cite{Wilkins2014,Bravin2012,Hanke2016,Momose2020,Tao2021,Birnbacher2021,Muller2022}. 
Notable laboratory-based XPC$\mu$T applications include: imaging of soft-tissue specimens like lung \cite{Velroyen2015,Murrie2020}, breast \cite{Massimi2021}, heart \cite{Reichardt2020}, oesophagus \cite{Savvidis2022} and brain \cite{Topperwien2017}, multi-material phantoms \cite{Willner2016,Weber2015,Ji2020} and for composites like carbon fibre-reinforced composite materials \cite{Shoukroun2020}. 

In this context, we have identified a solution for lab-XPC$\mu$T that has minimum requirements on the experimental set-up. Our approach is based on a low-power, fixed target and sealed X-ray tube, it requires a single optical element (modulator) and it is compatible with flat-panel detectors as well as with more sophisticated single-photon counters. All of these components are readily available and require a minimal amount of maintenance. The modulator is fabricated through laser ablation of a tungsten foil and does not need to be positioned with high accuracy or repeatability. The total length of the system is less than one meter, making it one of the most compact for lab-XPC$\mu$T. Our approach shares similarities with a Shack-Hartmann wavefront sensor and in our work it evolved from one-directional to two-directional sensitivity of beam tracking \cite{vittoria2015,dreier2020}, it can also be viewed as a highly parallelised version of a single-probe scanning system \cite{deJonge2008}. We call this approach two-directional beam tracking (2DBT), this solution offers a compact, robust and cost-effective way of producing multi-contrast three-dimensional images of the inner structure of samples, without compromising on sensitivity.

\section*{Materials and methods}
\begin{figure*}[t]
\centering
\includegraphics[width=\textwidth]{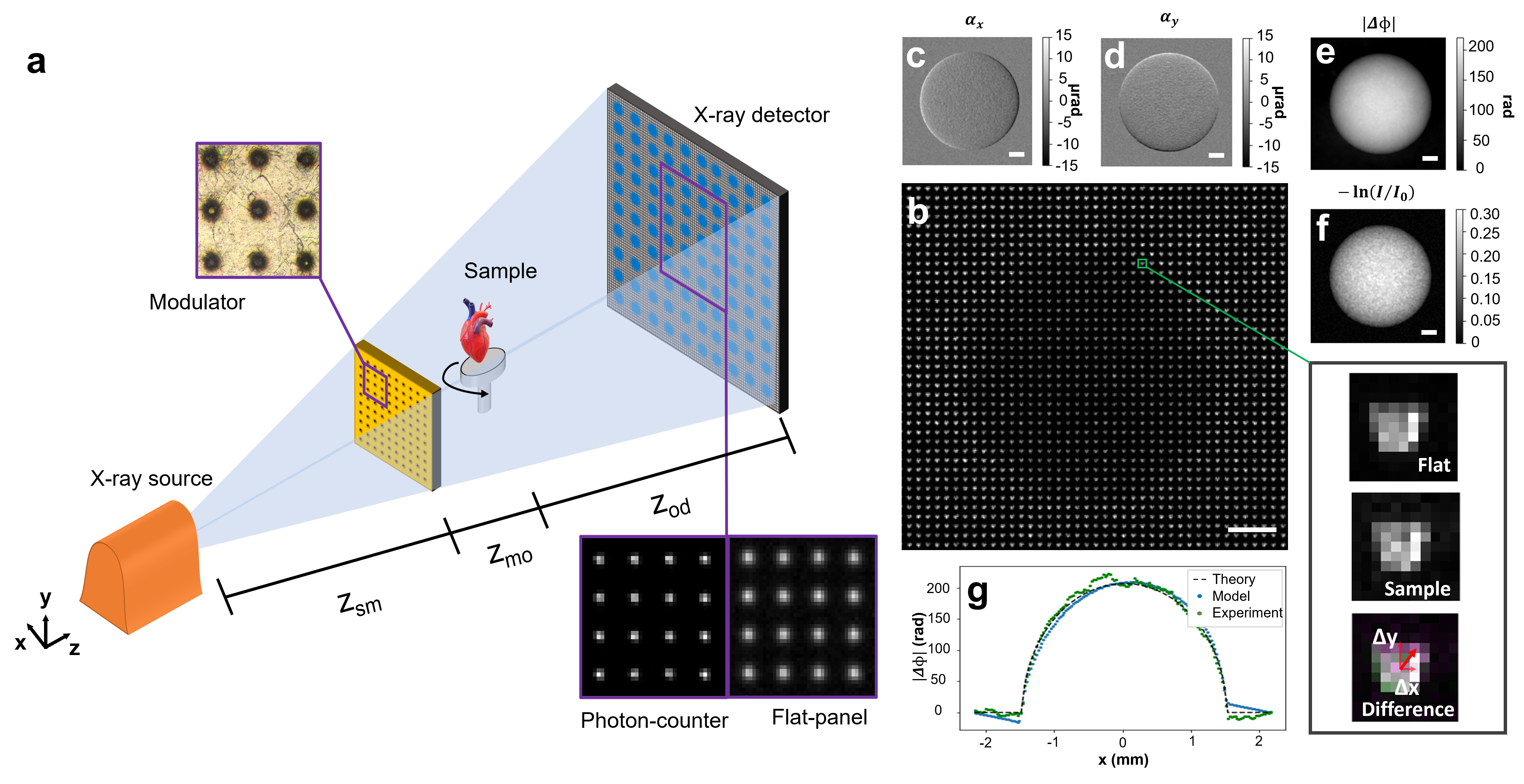}
\caption{\textbf{System's set-up and working principle}. (a) A schematic of the laboratory set-up, including a sealed microfocus X-ray tube, the modulator, sample stage and X-ray detector. An experimental image of a PMMA calibration sphere with an inset comparing beamlets without and with the sample in place (b). The effects of the presence of the sample are barely detectable by eye inspection, however the dedicated data analysis algorithm reliably extracts refraction images (c, d). These are subsequently integrated leading to phase images (e) that offer superior signal-to-noise ratio to the standard attenuation-contrast image (f). The refraction images are retrieved with a sub-pixel cross-correlation algorithm, we observe good agreement between the phase shift expected from theory, obtained through the numerical simulation and from experimental data (f). All scale bars are $500$ $\mu$m.
}
\label{fig1:system}
\end{figure*}

\subsection*{Model}
The core element of the imaging system (Figure \ref{fig1:system}a) is the structured illumination, obtained by spatially modulating the amplitude of the radiation field before it reaches the sample. Structuring the illumination in two dimensions enables tracking small changes in the position of each probe, i.e. measuring refraction along two orthogonal directions. It redefines sampling and spatial resolution of the imaging system. Sampling is no longer governed by the detector pixel pitch, it is instead defined by the distance between apertures in the modulator. The spatial resolution is redefined as being equal to or better than the width of the apertures in the modulator \cite{Diemoz2014, Hagen2018}. The main drawback of this approach is the sacrifice in flux available for imaging. We note, however, that this sacrifice does not affect dose efficiency because the radiation field is shaped before it reaches the sample. We also note that sampling is not necessarily limited by the pitch of the modulator: finer sampling can be achieved by recombining subsequent exposures, having moved either the modulator or the sample by a fraction of the pitch. We increase sampling by displacing the modulator by equally spaced sub-pitch increments. This requires a reference image for each modulator position, however it preserves the conventional cone beam geometry and allows using standard 3D reconstruction algorithms, even for the data sets acquired with higher sampling. Each probe is detected, and subsequently analysed, independently. The X-ray intensity distribution of the probe is modelled through a series of convolutions:
\begin{equation}
    \label{eq:model}
    I_0(x,y)=(S*M*PSF)(x,y)
\end{equation}
where $*$ denotes the two-dimensional convolution operator, $S$ is the source intensity distribution, geometrically scaled by $-1 + (z_{sm} + z_{mo} + z_{od})/z_{sm}$, $M$ the transmission of the modulator (unity over a circle representing the aperture and zero elsewhere) geometrically scaled by $(z_{sm} + z_{mo} + z_{od})/z_{sm}$, and $PSF$ the detector point spread function. Referring to Figure \ref{fig1:system}a, $z_{sm}$ is the distance between the source and the modulator, $z_{mo}$ is the distance between the modulator and the axis of rotation and $z_{od}$ is the distance between the axis of rotation and the detector.

When a sample is in place, the intensity distribution of the probes is attenuated and shifted by the local properties of the sample:
\begin{equation}
    I(x,y)=tI_0(x-\Delta x, y-\Delta y)
\end{equation}
where $t$ is the transmission through the sample (at the position of the probe) and $\Delta x$ and $\Delta y$ are the probe displacements on the detector plane due to the refraction induced by the sample. These shifts are measured by comparing the probe intensity distributions with and without the object (Figure \ref{fig1:system}b). The transmission is calculated as the ratio of the total intensity in the probe:
\begin{equation}
    \label{eq:tranmission}
    t=\frac{\sum I(x,y)}{\sum I_0(x,y)},
\end{equation}
where sum operates over all the pixels illuminated by a single probe (Figure \ref{fig1:system}b, zoom-ins). For the retrieval of the displacements $\Delta x$ and $\Delta y$ we used a sub-pixel cross-correlation algorithm \cite{Guizar2008}. These displacements are related to the refraction angle and to the object-to-detector distance, $z_{od}$, by:
\begin{equation}
    \alpha_{x}(x, y) = \tan^{-1}\left(\frac{\Delta x}{z_{od}}\right) \;,\;
    \alpha_{y}(x, y) = \tan^{-1}\left(\frac{\Delta y}{z_{od}}\right).
\end{equation}
By linking the two refraction images to the orthogonal gradients of the phase shift (Figure \ref{fig1:system}, panels c and d) we obtain:
\begin{equation}
    \alpha_x(x, y) = \frac{1}{k}\Delta \Phi_x(x, y)
    \;,\;
    \alpha_y(x, y) = \frac{1}{k}\Delta \Phi_y(x, y)
\end{equation}

where $k$ is the wave-number. The phase-shift of the wavefront $\Delta \Phi(x, y)$ (Figure \ref{fig1:system}e) can be obtained by means of the Fourier derivative theorem \cite{Kottler2007}:
\begin{equation}
   \Delta \Phi(x, y) = \mathcal{F}^{-1}\left[\frac{\mathcal{F}[\Delta \Phi_x+i \Delta \Phi_y](u,v)}{2\pi i(u+iv)}\right](x,y)
    \label{eq:fourier_int}
\end{equation}
where $(u,v)$ represent the reciprocal space coordinates to $(x, y)$. A better signal-to-noise ratio can be seen by direct visual comparison with the attenuation-image contrast of the same phantom (Figure \ref{fig1:system}f). 
The retrieved quantities $t$ and $\Delta \Phi$ are line integrals along the photon path of two physical properties of the sample, the linear attenuation coefficient ($\mu$) and the real part of the refractive index ($\delta$):
\begin{equation}
    \label{eq:mu}
    -\ln t(x,y) = \int_o\mu(x',y',z') dz
\end{equation}

\begin{equation}
    \label{eq:delta}
    -\frac{\Delta \Phi (x,y)}{k} = \int_o\delta(x',y',z')dz
\end{equation}
these relationships allow one to obtain volumetric reconstructions of $\mu$ and $\delta$ from projections taken at different viewing angles using the standard algorithms for tomography.

\subsection*{Experiments}
The X-ray system features a fixed W-target microfocus Hamamatsu X-ray source (L12161-07) which we run at 40 kVp and 10 W. The modulators were manufactured with a laser ablation procedure from readily available 100 $\mu$m-thick Tungsten foils (Goodfellow). Two pitches, 50 and 100 $\mu$m, were tested. The apertures have a conical shape with diameters of $\sim$15 $\mu$m and $\sim$30 $\mu$m in the front and back apertures, respectively, the narrower aperture was facing the source.

Three detectors were tested: i) Pixirad-2/PIXIE-III photon-counter with a 650 $\mu$m CdTe sensor (98\% detection efficiency up to 50 keV), 62 $\mu$m pixel pitch and a $50\times32$ mm$^2$ active area; ii) MerlinX photon-counter (Quantum Detectors) with a Medipix 3RX chip (256x256), a 500 $\mu$m Si sensor and 55 $\mu$m pixel pitch; and iii) Hamamatsu Flat Panel (C9732DK-11) with directly deposited CsI scintillator, 50 $\mu$m pixel pitch and $12\times12$ cm$^2$ active area.
The geometry was defined as follows: $z_{sm} = (150, 150, 140)$ mm, $z_{mo} = (30, 30, 25)$ mm and $z_{od} = (680, 580, 535)$ mm, for the Pixirad, MerlinX and Hamamatsu detectors, respectively. Single-photon-counters are preferable when sensitivity is the priority, while flat panels offer a much larger field of view albeit with lower sensitivity.

To characterise the system and fine-tune the quantitative retrieval algorithm, a test phantom consisting of a Poly(methyl methacrylate) (PMMA) calibration sphere (Goodfellow, $\diameter = 3.18 \pm 0.05$mm) was used. For the experimental assessment of spatial resolution we used a phantom composed of soda-lime glass micro-spheres (Fischer Scientific, monodisperse, 50 $\mu$m diameter) deposited on a kapton substrate. The detector was the Pixirad with the 100 $\mu$m-pitch modulator and noise threshold set at 10 keV. A total of $16\times16$ frames were acquired for both phantoms, with a sampling step in both $x$ and $y$ directions of $6.25$ $\mu$m and 1 s exposure time per frame. 
 
The angular sensitivity was quantified on planar images of a phantom composed of soda-lime glass micro-spheres (Fischer Scientific, monodisperse, 50 $\mu$m diameter) embedded in wax and polyethylene foam. Images were acquired at $4\times4$ positions, with $x$ and $y$ steps of $12.5$ $\mu$m. The detector was the MerlinX with the 50 $\mu$m-pitch modulator and noise threshold set at 5 keV. For each step 256 frames of 0.25 s exposure time were acquired to study the sensitivity as a function of exposure time. The angular sensitivity was measured in an area without the sample by calculating the mean and standard error of the standard deviation of the measured refraction angles \cite{Modregger2011}, 9 different windows of $10\times10$ pixels each were used.

The ability of the system to measure $\mu$ and $\delta$ in three dimensions was assessed with a multi-material phantom composed of three spheres (Goodfellow) in a plastic straw. The materials were Polypropylene (PP),  Polystyrene (PS) and PMMA with diameters: $\diameter_{PP, PMMA}= 3.18 \pm 0.05$ mm, $\diameter_{PS} = 3.5 \pm 0.1$ mm.  The detector was the Pixirad with the 100 $\mu$m-pitch modulator and noise threshold set at 8 keV. We acquired 900 projections over 360$^{\circ}$ with 1 s exposure time per projection. A full rotation was acquired for each modulator position, giving a total of $8\times8$ scans with sampling every 12.5 $\mu$m in $x$ and $y$. 

The performance of the system for soft-tissue imaging was demonstrated with two animal-derived small organs; a heart and an oesophagus, extracted from a 300 g Sprague-Dawley rat and a 3 kg piglet respectively. The heart sample was obtained from the UCL BSU, whilst the oesophagus from piglets procured by JSR genomics, a Home Office approved supplier. Animals were euthanised via Schedule 1 methods and sample sizing has been implemented following the NC3R principles. The organs were prepared for imaging through paraformaldehyde (PFA) fixation, followed by critical point drying (as described in previous work \cite{Savvidis2022}). For the piglet oesophagus we acquired 900 equally spaced projections over 360$^{\circ}$ with 1 s exposure time using the Pixirad detector. The $100 \, \mu$m-pitch modulator was stepped by 12.5 $\mu$m in $x$ and $y$ for a total of $8\times8$ scans and an exposure time of $900\times8\times8\times1$ s $=16$ h. For rat heart we acquired 2000 equally spaced projections over 360$^{\circ}$ with 1.2 s exposure time using the Hamamatsu flat-panel. The $50 \, \mu$m-pitch modulator was stepped $4\times4$ times on a grid with in $12.5$ $\mu$m displacements in $x$ and $y$. The total exposure time was $2000\times4\times4\times1.2$ s $= 10.7$ h.

All reconstructions were performed using the Feldkamp-Davis-Kress algorithm implementation in CUDA of the Astra Toolbox \cite{astra_2015}. The volume renderings presented in Figure \ref{fig3:applications} were obtained using the Avizo software (Thermo Fischer Scientific).

\begin{figure*}[t]
\centering
\includegraphics[width=\textwidth]{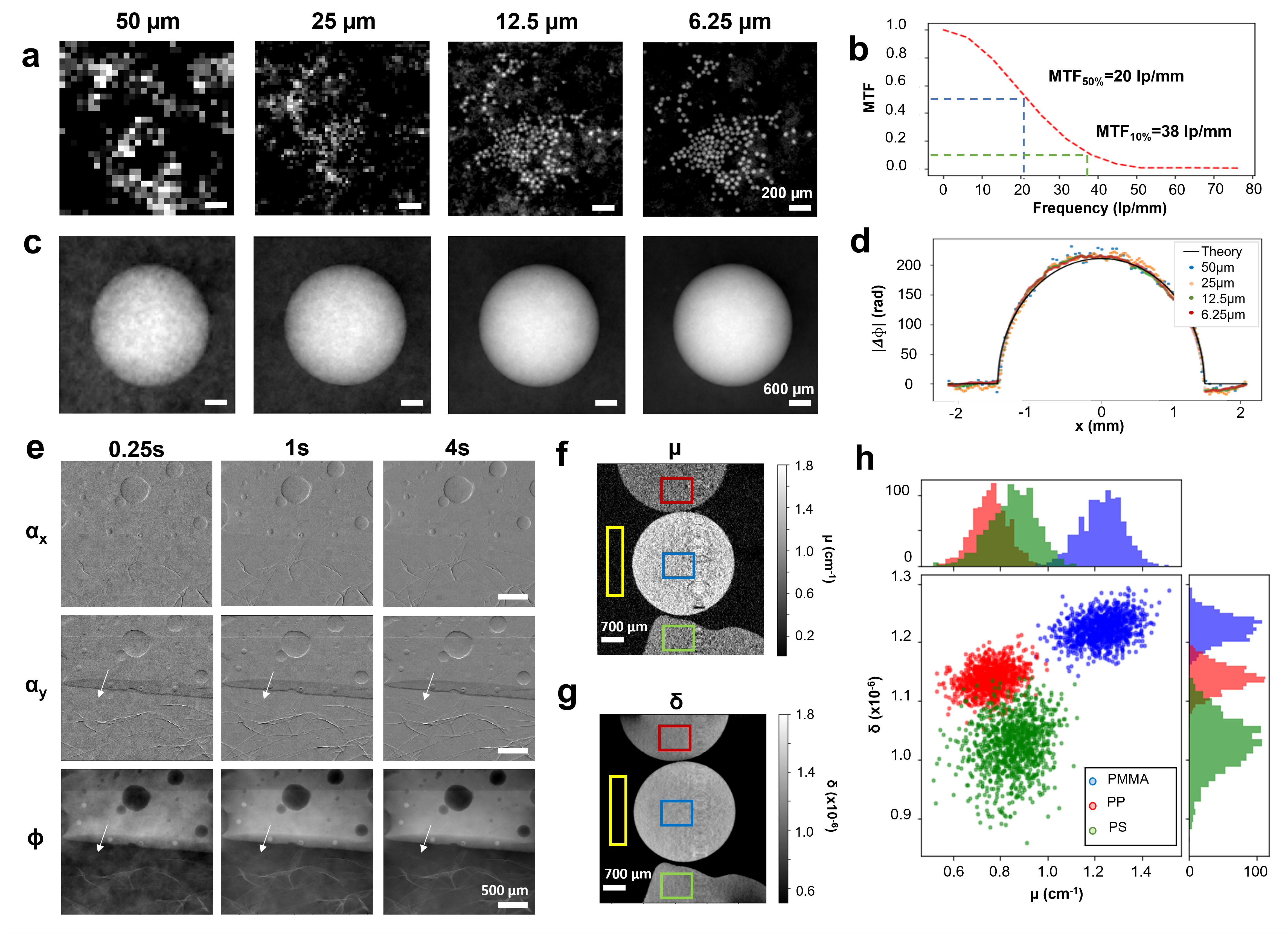}
\caption{\textbf{System characterisation and benchmarking}. (a) Resolution and (c) accuracy of phase retrieval, as a function of sampling, in images of the monodisperse soda-lime glass micro-spheres on Kapton substrate and the PMMA calibration sphere, respectively. (a) The ability to visualise small details depends critically on a fine sampling and is mainly defined by the width aperture in the modulator. The MTF (b) matches well a real-space full width half maximum of $16$ $\mu$m. (c) Phase is retrieved accurately in all cases, with higher noise linked to coarser sampling. This can also be seen in the line profiles plot (d) where the phase shifts retrieved in the four experimental cases are compared to the theoretical phase shift. (e) Images of the angular sensitivity phantom: the system sensitivity depends on the exposure time, and longer exposure times lead to better signal to noise ratios. Exposures as short as 250 ms are sufficient for robust phase integration. Longitudinal slices of (f) $\mu$ and (g) $\delta$ for the multi-material phantom. (h) The ability to separate the three materials is greatly improved by the simultaneous availability of the two contrast channels.}
\label{fig2:characterization}
\end{figure*}
\begin{figure*}[!ht]
\begin{floatrow}
\capbtabbox{
   \begin{tabular}{c|cc}
   \hline
       Exp time (s) & $\sigma(\alpha_x)$ $(\mu rad)$ & $\sigma(\alpha_y)$ $(\mu rad)$ \\ \hline
       0.25 & 2.14 $\pm$ 0.05 & 2.33 $\pm$ 0.07  \\ 
       0.5 & 1.62 $\pm$ 0.04 & 1.66 $\pm$ 0.03  \\ 
       1 & 1.10 $\pm$ 0.02 & 1.19 $\pm$ 0.03  \\ 
       2 & 0.82 $\pm$ 0.02 & 0.85 $\pm$ 0.02  \\ 
       4 & 0.61 $\pm$ 0.01 & 0.65 $\pm$ 0.01  \\ 
       8 & 0.43 $\pm$ 0.01 & 0.46 $\pm$ 0.01  \\ 
       16 & 0.35 $\pm$ 0.01 & 0.34 $\pm$ 0.01  \\ 
       32 & 0.27 $\pm$ 0.01 & 0.31 $\pm$ 0.01  \\ 
       64 & 0.22 $\pm$ 0.01 & 0.22 $\pm$ 0.01  \\ \hline
   \end{tabular}
}{
   \caption{\label{tab1:sensitivity}System sensitivity for different exposure times, MerlinX detector and 50 $\mu$m-pitch modulator.}
}
\ffigbox{
  \includegraphics[width=0.4\textwidth]{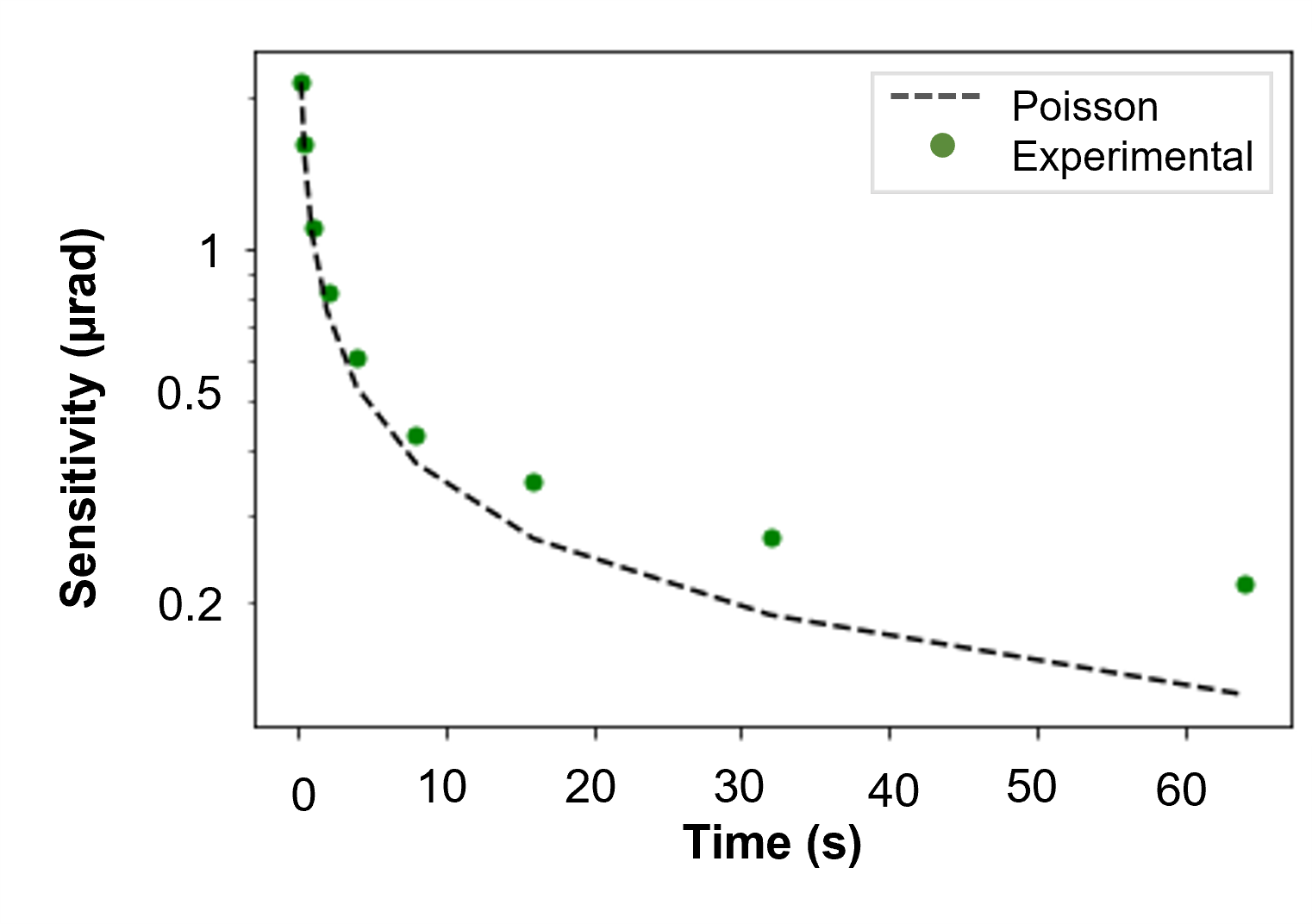}
}{
  \caption{\label{fig3:sensitivity}System sensitivity trend compared to quantum-limited ideal system, MerlinX detector and 50 $\mu$m-pitch modulator.}
}
\end{floatrow}
\end{figure*}

\section*{Results and discussion}
Figure \ref{fig2:characterization}a and \ref{fig2:characterization}c report the resolution phantom and the PMMA calibration sphere at different sampling steps ranging from $50$ to $6.25$ $\mu$m. The positioning of the modulator has to be accurate and repeatable in comparison to these numbers, namely approximately half of the aperture width at best, which is compatible with most micro-positioning technologies. Whilst the phase shift of a large object is correctly retrieved in all cases, the ability to represent finer details and distinguish between adjacent micro-spheres critically depends on the sampling step. The modulation transfer function of the system at $6.25$ $\mu$m sampling pitch is shown in \ref{fig2:characterization}b, with 50\% at $20$ lp/mm and 10\% at 38 lp/mm. It matches well with the expectation from a real-space full-width-half-maximum of $16$ $\mu$m, comparable to the aperture size in the modulator. Figure \ref{fig2:characterization}d shows four line plots extracted from the phase images in panel \ref{fig2:characterization}d against the theoretically expected phase shift. There is good agreement between the phase shifts measured with different sampling steps and theory, with higher noise linked to lower sampling. The decrease in noise that is linked to progressively finer sampling is also visible by visual comparison of the four images in panel \ref{fig2:characterization}c.

The angular sensitivity depends on the exposure time and closely follows the trend expected from the Poisson statistics (Table \ref{tab1:sensitivity} and Figure ). This indicates that the main limitation in the phase retrieval process is photon statistics. The gain in sensitivity with exposure time is slightly worse than in a perfect quantum-limited system, this is due to environmental instabilities including temperature drifts and vibrations, but these effects become noticeable only with tens of seconds of integration. The two differential phase contrast images and the integrated phase of the sensitivity phantom are presented in Figure \ref{fig2:characterization}e. Smaller details emerge from the background noise as the exposure time is increased. We report sub-micro-radian sensitivities starting at 2 s of exposure time and down to minimum of 220 $\pm$ 10 $n$rad with 64 s per frame. These values are obtained with a single exposure image, thus with a resolution limited by the modulator pitch at 50 $\mu$m. With the methods described earlier, the resolution can be increased up to approximately $16$ $\mu$m while maintaining the same angular sensitivity and increasing the total exposure time. The sensitivity with the Hamamatsu flat-panel was slightly worse, ($\sigma(\alpha_x)$, $\sigma(\alpha_y)$) = (1.96 $\pm$ 0.05, 2.10 $\pm$ 0.05) $\mu$rad and ($\sigma(\alpha_x)$, $\sigma(\alpha_y)$) = (1.31 $\pm$ 0.04, 1.41 $\pm$ 0.02) $\mu$rad for 1 s and 2 s exposure time, respectively. However, the field of view was extended to 28 mm $\times$  28 mm from the 3.2 mm $\times$  3.2 mm achievable with the MerlinX. To position our system in the landscape of existing solutions for lab-based XPC$\mu$T we report in Table \ref{tab2:comparison} a summary of imaging system parameters from the literature, including source power, spatial resolution, system dimensions and sensitivity.
\begin{table*}[ht]
    \centering
    \begin{adjustbox}{width=1\textwidth}
    \begin{tabular}{lcrrrrrr}
    \hline
        Reference & Technique & Resolution & Sensitivity & Scan time (s) & X-ray spot size & Source power & Dimension \\ \hline \hline
        Diemoz el al \cite{Diemoz2013} & EI & 12 or 66.8 $\mu$m & 270 nrad  & 112 or 14s & 70 $\mu$m & 875W & 200 cm\\
        Havariyoun el al \cite{Havariyoun2019} & EI & 10 or 79 $\mu$m & 230 nrad & 384 or 24s & 70 $\mu$m & 400W & 100 cm\\ 
        Revol et al \cite{Revol2010} & GBI & 104 $\mu$m & 110 nrad & 80.4s & 1x1 mm$^2$ & 250-1000W & 140 cm \\
        Thuring et al \cite{thuring2013} & GBI & 4.1-7.1 $\mu$m & 250-550 nrad & 128s & 5-10 $\mu$m & 4-12W & 20-33 cm \\
        Birnbacher et al \cite{Birnbacher2016} & GBI & 170 $\mu$m & 5 nrad & 275s & 132x226 $\mu$m$^2$ & 2.8kW & 230 cm \\
        Villa-Comamala et al \cite{VilaComamala2021} & GBI & 21.5 $\mu$m & 45 nrad & 250s & 10 $\mu$m & 60W & 90.4 cm \\ 
        Zanette et al \cite{Zanette2014} & SBI & 84 $\mu$m & 240 nrad & 300s & 7.8x8.7 $\mu$m$^2$ & 30W & 300 cm\\
        Quenot et al \cite{Quenot2021} & SBI & 36 $\mu$m & 5 $\mu$rad & 600s & 4 $\mu$m & - & 53 cm\\
        Our set-up & 2DBT & 15 or 50 $\mu$m & 345 nrad & 256 or 16s & 5-20 $\mu$m & 10W & 70-86 cm\\\hline
    \end{tabular}
    \end{adjustbox}
    \caption{Comparison of laboratory-based XPC$\mu$T systems in literature}\label{tab2:comparison}
\end{table*}
In comparison with EI and the first GBI systems, our approach achieves comparable sensitivity with a low-power source. The highest sensitivity was achieved with GBI \cite{Birnbacher2016} and the most powerful X-ray source, at the lowest spatial resolution. A more recent lab-based implementation of GBI \cite{VilaComamala2021} achieved high sensitivity at high resolution, and required fine-pitch gratings with high aspect ratio. SBI \cite{Zanette2014} also reports high sensitivity and requires a large footprint and specialised source technology.

Tomographic reconstructions of $\mu$ and $\delta$ of the multi-material phantom are presented in Figure \ref{fig2:characterization}f and \ref{fig2:characterization}g. The lower noise of phase tomography is apparent. This can be seen quantitatively in the histograms in panel \ref{fig2:characterization}h where the values extracted from the boxes highlighted in colours are compared. Phase contrast allows for a better separation of PP and PS with respect to attenuation contrast, and the simultaneous availability of both contrast channels improves material separation even further. Measured values are compared against theoretically expected ones \cite{henke_1993} in Table \ref{tab2:quantitative}. Effective energies vary slightly for $\delta$ and $\mu$ as expected \cite{Munro2013}, however the range of values is very close to the mean energy of the X-ray spectrum ($15.6$ keV). The contrast-to-noise ration is calculated as:
\begin{equation}
    \label{eq:CNR}
    CNR = \frac{\mu_m - \mu_{air}}{\sigma_{air}} 
\end{equation}
where $\mu_m$ is the linear attenuation coefficient of each material and $\sigma_{air}$ is the standard deviation of the background. An analogous formula is used for $\delta$. CNRs were found to be substantially higher for phase tomography.

\begin{figure*}[h!]
\centering
\includegraphics[width=\textwidth]{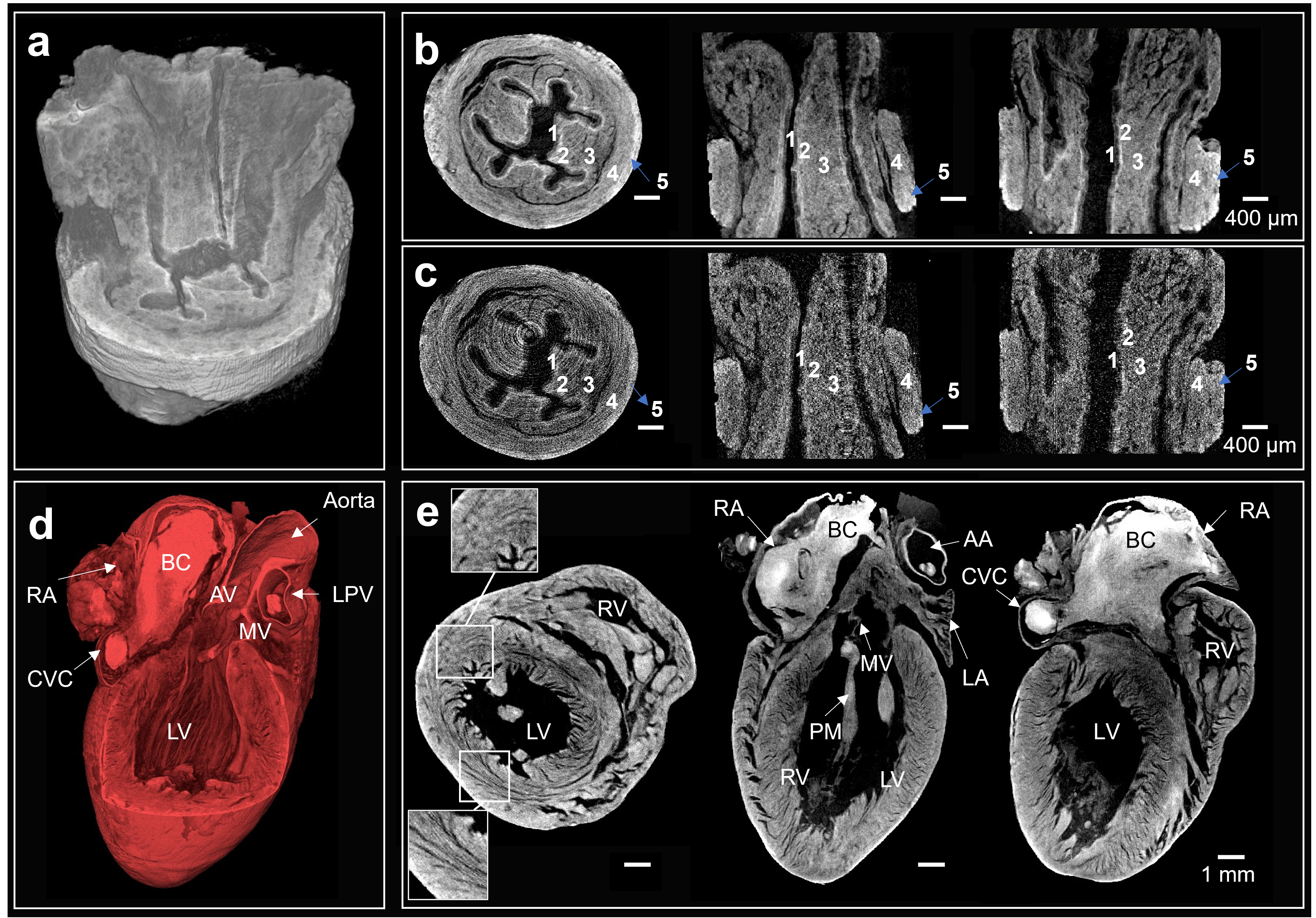}
\caption{\textbf{X-ray phase-contrast micro tomography of a piglet esophagus (a-c) and a rat heart (d, e)}. Piglet esophagus: (a) volume rendering and (b) phase contrast micro tomography in the axial, longitudinal and coronal slices (from left to right) and same slices for (c) conventional attenuation-contrast micro tomography. The numerical labels correspond to five different tissue layers that were identified in the phase contrast channel: epithelium (1), lamina propria (2), submucosa (3), inner circular muscular layer (4) and longitudinal outer muscular layer (5). These layers are hardly if at all visible in the images of panel (c). For the rat heart: (d) volume rendering of the contrast micro tomography (e) and its axial, longitudinal and coronal slices (from left to right). The following structures were identified: the left pulmonary vein (LPV), the aorta, the aortic valve (AV), the aortic arch (AA), the mitral valve (MV), the left ventricle (LV), the right ventricle (RV), the papillary muscles (PM), the right atrium (RA), and the caudal vena cava (CVC). We note the presence of a blood clot (BC) within the RA. The circumferentially oriented fibres in the mid-miocardium and the longitudinally oriented fibres closer to the outer epicardium wall can also be observed, as highlighted by the two zoom-in insets in panel (e).}
\label{fig3:applications}
\end{figure*}

\begin{table*}[ht]
\centering
\begin{adjustbox}{width=1\textwidth}
\begin{tabular}{|c|cccc|cccc|}
\hline
Material  & $\mu_{exp}$ (cm$^{-1}$) & $\mu_{theo}$ (cm$^{-1}$) & $E_\mu$ (keV) & CNR of $\mu$ & $\delta_{exp}$ (x10$^{-6}$) &  $\delta_{theo}$ (x10$^{-6}$) & $E_\delta$ (keV) & CNR of $\delta$\\
\hline
PP & 0.76 $\pm$ 0.06  & 0.76   & 13  & 4.1 & 1.14 $\pm$ 0.01 & 1.14 & 14 & 31.6\\
PMMA   &  1.23 $\pm$ 0.06 & 1.23 & 15  & 7.1 & 1.22 $\pm$ 0.01 & 1.22 & 15 & 34.0\\
PS   &  0.86 $\pm$ 0.06 &  0.85   & 14 & 4.7 & 1.03 $\pm$ 0.01 & 1.03 & 15 & 28.4\\
\hline
\end{tabular}
\end{adjustbox}
\caption{Experimental and theoretical $\mu$ and $\delta$ of the phantom materials, along with effective energies and CNRs measured with respect to the background (yellow box in Figure \ref{fig2:characterization}g). Good agreement is observed on both attenuation- and phase-contrast channels with expectations from theory. Phase tomography offers consistently higher CNRs.}
\label{tab2:quantitative}
\end{table*}
Lab-XPC$\mu$T of the animal-derived small organs are presented in Figure \ref{fig3:applications}. The volume renderings and coronal, longitudinal and axial slices of both samples are shown. Access to the full volume provides structural information of the specimens which are represented with isotropic spatial resolution of approximately $16$ $\mu$m.

For the piglet oesophagus (Figure \ref{fig3:applications}a-c), we identified five different tissue layers in the phase reconstructions as labelled in the images: epithelium (1), lamina propria (2), submucosa (3), inner circular muscular layer (4) and longitudinal outer muscular layer (5). These tissue layers are poorly, if at all, visible in the attenuation-contrast reconstructions. 
For the rat heart (Figure \ref{fig3:applications}d and \ref{fig3:applications}e), lab-XPC$\mu$T allowed to visualise the orientation of micro-fibres in the muscle bundles and to identify the following structures: the left pulmonary vein (LPV), the aorta, the aortic valve (AV), the aortic arch (AA), the mitral valve (MV), the left ventricle (LV), the right ventricle (RV), the papillary muscles (PM), the right atrium (RA), and the caudal vena cava (CVC). We note the presence of an iron-rich blood clot (BC) appearing brighter within the RA. The insets in the axial slice show the circumferentially oriented fibres in the mid-miocardium (line-like structures) and the longitudinally oriented fibres closer to the outer epicardium wall (point-like structures). 
The tomography datasets were acquired in a step-and-shoot fashion which carried overheads. Without optimisation the total acquisition times were in the range of approximately 64 h, this mismatch is only due to a need for optimising the sequences of positioning and detector exposure. Work is currently underway to implement fly-scans for bringing the data acquisition time much closer to the actual exposure times.

\section*{Conclusion}
We presented a compact and cost-effective system for performing X-ray phase-contrast micro tomography in a laboratory setting. The system is based on a single modulator, a low-power sealed X-ray tube and is compatible with a range of readily available detectors. We report sub-micron angular sensitivity starting at 2 seconds of exposure time at $50$ $\mu$m spatial resolution and 32 seconds at $16$ $\mu$m spatial resolution, and a maximum angular sensitivity of 220 nrad in $64$ s at $50$ $\mu$m spatial resolution. These results are compatible with the state of the art and were achieved with a simple and compact set-up. The proposed approach is quantitative, as validated through a phantom composed of known materials, and flexible in terms of spatial resolution and sampling. By introducing the amplitude modulator, sampling and resolution are driven by the structure of the illumination and can be tuned to the requirement of the sample by adapting the data acquisition strategy. The potential of the system for soft tissue imaging was demonstrated on two biological specimens prepared without staining: five different tissue layers were identified in a piglet oesophagus and the orientation of micro-fibers within the myocardium of a rat heart was visualised. We believe that the concurrent simplicity, sensitivity, robustness, compactness and efficiency of our approach will be instrumental in making X-ray phase-contrast micro tomography more accessible and available to a wider community.

\section*{Author contributions}
Conceptualization ME; 
Methodology ME CN-L AD GP SC AA;
Software CN-L AD AA CKH;
Investigation CN-L AD ME; 
Resources ME GP SC AA SS MFMG AO; 
Writing CN-L ME; 
Review CN-L AD SS MFMG GP AA SC DB CKH AO ME; 
Supervision AO DB CKH SC ME; 
Funding acquisition AO DB CKH ME. 

\section*{Acknowledgements}
This work was supported by the EPSRC EP/T005408/1; Wellcome Trust 221367/Z/20/Z; National Research Facility for Lab X-ray CT (NXCT) through EPSRC grants EP/T02593X/1 and EP/V035932/1. AO is supported by the Royal Academy of Engineering under their Chair in Emerging Technologies scheme (CiET1819/2/78). CKH is supported by the Royal Academy of Engineering under the Research Fellowship Scheme. MFMG is supported by an H2020 Marie Skłowdoska Curie Action Fellowship (AmnioticID, grant agreement: 843265). SS is a UKRI EPSRC Doctoral Prize Fellow (EP/T517793/1).

\bibliographystyle{ieeetr}

\end{document}